\documentclass{article}
\usepackage{spconf,amsmath,graphicx}
\usepackage{stmaryrd}
\usepackage{bm}
\usepackage{amssymb}
\usepackage{booktabs}
\usepackage{graphicx}
\usepackage{float}
\usepackage{graphicx}
\usepackage{stfloats}
\usepackage{amsmath}
\usepackage{epstopdf}
\usepackage{algorithm,algorithmic}
\usepackage{multirow}
\usepackage{bbm}
\usepackage{framed}
\usepackage[colorlinks,linkcolor=red,anchorcolor=green,citecolor=green]{hyperref}
\usepackage[numbers,sort&compress]{natbib}
\usepackage{amsthm}

\newcommand{\ie}{{\em i.e.}}
\newcommand{\eg}{{\em e.g.}}

\DeclareMathOperator*{\argmin}{arg\,min}

\title{The Power of Triply Complementary Priors for Image Compressive Sensing}
\name{Zhiyuan~Zha$^{1}$, Xin~Yuan$^{2}$, Joey Tianyi Zhou$^{3}$, Jiantao~Zhou$^{4}$, Bihan~Wen$^{1*}$\thanks{* Corresponding Author.} and Ce~Zhu$^{5}$}
\address{$^1${School of Electrical \& Electronic Engineering, Nanyang Technological University, Singapore 639798.}\\
$^2${Nokia Bell Labs, 600 Mountain Avenue, Murray Hill, NJ, 07974, USA.}\\
$^3${Institute of High Performance Computing, A*STAR, Singapore 138632.}\\
$^4${Department of Computer and Information Science,  University of Macau, Macau 999078, China.}\\
$^5${School of Information and Communication Engineering,}\\
{University of Electronic Science and Technology of China, Chengdu, 611731, China.}\\
}
\begin{document}
\ninept
\maketitle
\begin{abstract}
Recent works that utilized deep models have achieved superior results in various image restoration applications. Such approach is typically supervised which requires a corpus of training images with distribution similar to the images to be recovered. On the other hand, the shallow methods which are usually unsupervised remain promising performance in many inverse problems, \eg, image compressive sensing (CS), as they can effectively leverage non-local self-similarity  priors of natural images. However, most of such methods are patch-based leading to the restored images with various ringing artifacts due to naive patch aggregation. Using either approach alone usually limits performance and generalizability in image restoration tasks. In this paper, we propose a joint low-rank and deep (LRD) image model, which contains a pair of triply complementary priors, namely \textit{external} and \textit{internal}, \textit{deep} and \textit{shallow}, and \textit{local} and \textit{non-local} priors. We then propose a novel hybrid plug-and-play (H-PnP) framework based on the LRD model for image CS. To make the optimization tractable, a simple yet effective algorithm is proposed to solve the proposed H-PnP based image CS problem. Extensive experimental results demonstrate that the proposed H-PnP algorithm  significantly outperforms the state-of-the-art techniques for image CS recovery such as  SCSNet and WNNM.
\end{abstract}
\begin{keywords}
Image CS, triply complementary priors, hybrid plug-and-play, deep prior, non-local self-similarity.
\end{keywords}
\vspace{-2mm}
\section{Introduction}
Compressive sensing (CS) has attracted considerable interests for many researchers in signal and image processing communities \cite{candes2006robust,donoho2006compressed}. 
The most attractive aspect of CS is that the sampling and compression are conducted simultaneously, and almost all computational cost is derived from the decoder stage, and therefore, leading to a low computational cost in the encoder stage \cite{zhang2018ista}.
Due to the {unique advantages} of CS, it has been widely used in many important {applications}, such as magnetic resonance imaging \cite{wen2020transform} and single-pixel camera \cite{duarte2008single}.

\begin{figure}[!t]
\begin{minipage}[b]{1\linewidth}
{\includegraphics[width = 1\textwidth]{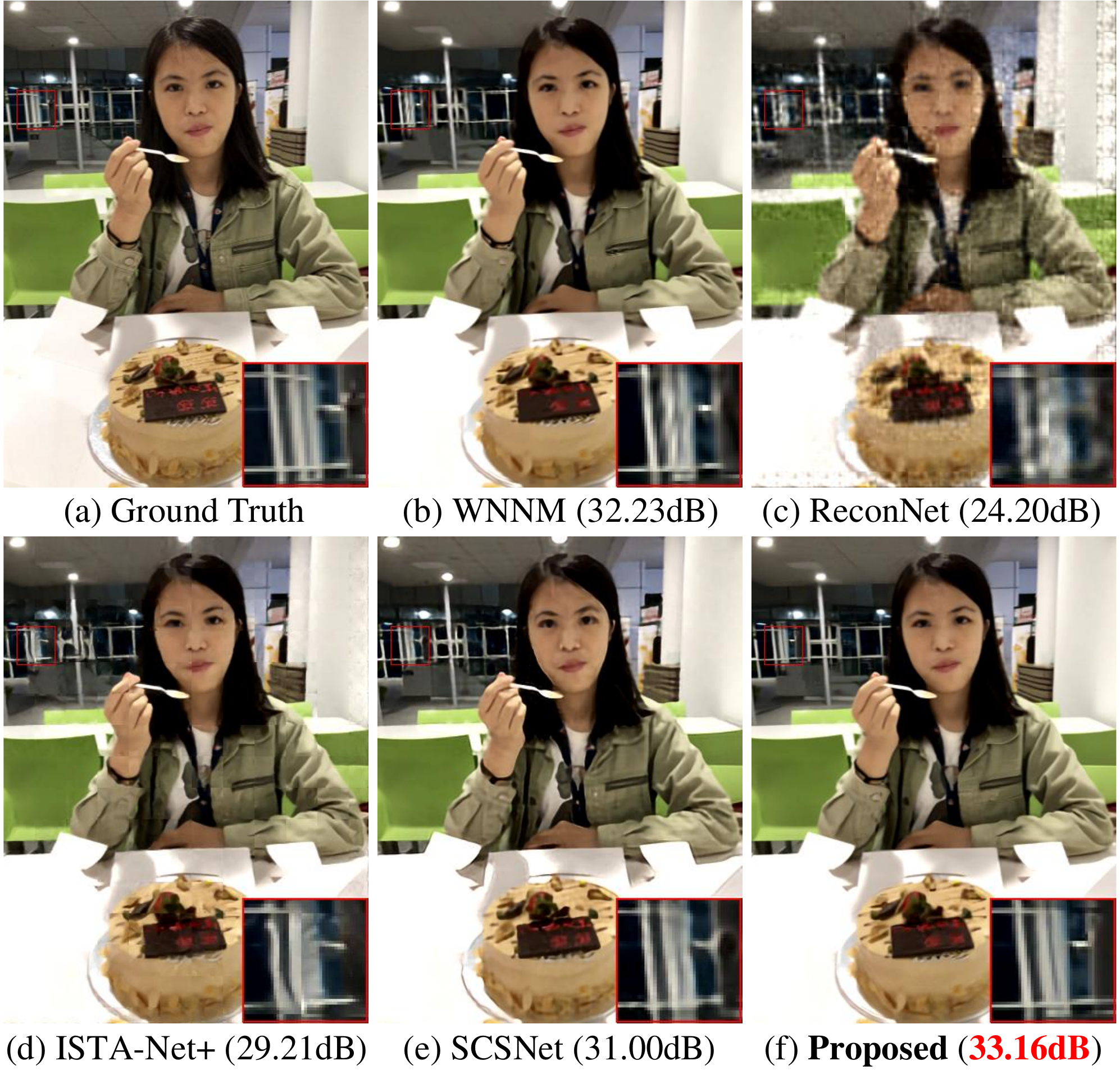}}
\end{minipage}
\vspace{-6mm}
\caption{\footnotesize CS results produced by our proposed H-PnP and state-of-the-art algorithms, when the CS ratio is 10\%. (a) The original image; (b) WNNM \cite{gu2017weighted};  (c) ReconNet \cite{kulkarni2016reconnet}; (d) ISTA-Net$^{+}$ \cite{zhang2018ista}; (e) SCSNet \cite{shi2019scalable}; (f) the restored image by our proposed H-PnP, where the ringing artifacts have been eliminated significantly  and sharp details are recovered.}
\label{fig:1}
\vspace{-6mm}
\end{figure}

Image CS methods can be classified, in general, into two categories: shallow model-based methods \cite{zhang2012image,li2009user,zhang2014group,zha2020benchmark,eslahi2016compressive,geng2018truncated} and deep learning based methods \cite{kulkarni2016reconnet,zhang2018ista,shi2019scalable,yuan2018parallel,mousavi2015deep,8765626}.
{One of the classic shallow approaches is based on image sparsity model, assuming that each image local patch can be encoded as a linear combination of basis elements \cite{elad2006image}.
Recent works exploited image non-local self-similarity (NSS)  prior \cite{mairal2009non}, by clustering similar patches into groups and modeling them by structural sparsity \cite{zhang2014group,mairal2009non,eslahi2016compressive} or low-rankness \cite{gu2017weighted,geng2018truncated,8931738}.
However, most of the shallow methods are patch-based, which inevitably lead to recovered images with ringing or blocky artifacts due to naive patch aggregation (See an example in Fig.~\ref{fig:1} (b)). 
}

{On the other hand, many recent works applied various deep neural networks for image restoration tasks \cite{zhang2017learning,kulkarni2016reconnet,zhang2018ista,shi2019scalable,yuan2018parallel,mousavi2015deep,8765626,zhang2018ffdnet}.
Such approaches have demonstrated great potential to learn image properties from training dataset with an end-to-end approach.
Popular neural networks structures, such as convolutional neural networks (CNN) \cite{zhang2017learning} and recurrent neural networks (RNN) \cite{zhang2018dynamic}, have been applied to image restoration tasks achieving state-of-the-art results.
Since most of the deep algorithms are supervised, in applications such as  remote sensing or biomedical imaging, the model trained on a standard image corpus fails when applied to specific modalities. 
Furthermore, most of the existing deep methods focused on exploiting image local properties (due to limited receptive field) while largely ignoring NSS, which limit their performance in many image restoration applications.
}

\begin{table}[!t]
\vspace{-4mm}
\caption{\footnotesize Comparison of the key attributes between the proposed H-PnP and other representative methods for image restoration.}
\resizebox{0.485\textwidth}{!}				
{
\footnotesize
\centering
\begin{tabular}{|c|c|c|c|c|c|c|c|c|c|c|c|c|c|c|}
\hline					
\multirow{2}{*}{\textbf{{Methods}}}&\multirow{2}{*}{\textbf{{Non-local}}}&\multirow{1}{*}{\textbf{{Shallow}}}
&\multirow{1}{*}{\textbf{{Deep}}}&\multirow{1}{*}{\textbf{{Internal}}}&\multirow{1}{*}{\textbf{{External}}}\\
&&{\textbf{Model}}&{\textbf{Model}}&{\textbf{Prior}}&{\textbf{Prior}}\\
\hline
\multirow{1}{*}{TV \cite{li2009user}}
& 	&		\checkmark       &		        &   \checkmark    &
\\
\hline
\multirow{1}{*}{GSR \cite{zhang2014group}}
&	    \checkmark    	&	\checkmark	& 	&	 	\checkmark &	
\\
\hline
\multirow{1}{*}{GMM \cite{yang2014compressive}}
&	       	&	\checkmark	& 	&	  &	\checkmark
\\
\hline
\multirow{1}{*}{WNNM \cite{gu2017weighted}}
&	    \checkmark    	&	\checkmark	& 	&	 	\checkmark &	
\\
\hline
\multirow{1}{*}{ReconNet \cite{kulkarni2016reconnet}}
&	           &	        	&	   \checkmark      	& 	&	 \checkmark
\\
\hline
\multirow{1}{*}{ISTA-Net$^{+}$ \cite{zhang2018ista}}
&	           &	        	&	   \checkmark      	& 	&	 \checkmark	
\\
\hline
\multirow{1}{*}{NLRN \cite{liu2018non}}
&	  \checkmark         &	        	&	   \checkmark      	& 	&	 \checkmark	
\\
\hline
\multirow{1}{*}{SCSNet \cite{shi2019scalable}}
&	           &	        	&	   \checkmark      	& 	&	 \checkmark	
\\
\hline
\multirow{1}{*}{\textbf{Proposed}}
&	\checkmark & \checkmark    	&\checkmark   	&	\checkmark 	&	\checkmark
\\
\hline
\end{tabular}
}
\label{Tab:1}
\vspace{-5mm}
\end{table}

Bearing the above concerns in mind, we propose a joint low-rank and deep (LRD) image model, which contains a pair of triply complementary priors, namely \textit{external} and \textit{internal}, \textit{deep} and \textit{shallow}, and \textit{local} and \textit{non-local} priors.
Based on the LRD model, we propose a novel hybrid plug-and-play (H-PnP) framework for highly effective image CS. To the best of our knowledge, this is the first work to jointly exploit both NSS and deep priors under a unified framework for image restoration. Table~\ref{Tab:1} depicts the proposed LRD-based H-PnP scheme and some representative of existing reconstruction algorithms with their key attributes.

The main contributions of this paper are summarized as follows:

\begin{list}{\labelitemi}{\leftmargin=20pt \topsep=0pt \parsep=-3pt}
    \item The proposed LRD model jointly exploits triply-complementary low-rank and deep denoiser priors, to take the advantages of NSS, scalable model richness, as well as good generalizability. In practice, LRD-based method significantly improves the visual quality of the CS reconstructed images (see an example in Fig.~\ref{fig:1} (f)).

    \item  We propose the H-PnP framework for image CS based on the novel LRD model. To make the optimization tractable, we propose an efficient yet effective algorithm by applying alternating minimizing.

    \item Extensive experimental results demonstrate that the proposed H-PnP based image CS algorithm has superior performance comparing to several popular or state-of-the-art image CS methods.
\end{list}



\vspace{-2mm}
\section{Related Works}
\label{sec:2}

In this section,  we give a brief introduction on CS, deep PnP model and low-rank image modeling.

{\bf Compressive Sensing:}
The goal of image CS is to reconstruct the high-quality image $\textbf{\emph{x}}\in{\mathbb R}^{N}$ from its undersampled (lower than Nyquist sampling rate \cite{donoho2006compressed}) measurements $\textbf{\emph{y}}\in{\mathbb R}^{M}$ obtained by
\begin{equation}
\textstyle{\textbf{\emph{y}} = \boldsymbol\Phi\textbf{\emph{x}} + \textbf{\emph{e}},}
\label{eq:1}
\end{equation}
where $\boldsymbol\Phi\in{\mathbb R}^{M\times N}$ denotes the random projection matrix, $M \ll N$, and $\textbf{\emph{e}}\in{\mathbb R}^{M}$ is the additive noise or measurement error. The image CS is an ill-posed inverse problem as the measurement $\textbf{\emph{y}}$ has much lower dimension than that of the underlying image. Therefore, an effective prior is key to a successful image CS algorithm \cite{li2009user,shi2019scalable,zhang2014group}.

{\bf Deep Plug-And-Play Methods:}
Recent works on the plug-and-play (PnP) framework \cite{venkatakrishnan2013plug,ulyanov2018deep,zhang2017learning} allowed applying the effective image denoiser to solve the general inverse problems, such as image deblurring \cite{zhang2017learning}, image inpainting \cite{8489894} and computational imaging  \cite{8464091}, etc. By decoupling the problem-specific sensing modality (\ie, the $\boldsymbol\Phi$ for image CS) from the general image priors, PnP provides a more flexible approach to generalize denoising algorithms to other more  sophisticated applications. Very recent works \cite{zhang2017learning,8464091,8489894,ulyanov2018deep} applied state-of-the-art deep denoisers in PnP by solving the following maximum a posteriori (MAP) problem:
\begin{equation}
\textstyle{\arg\min\limits_{\textbf{\emph{x}}}\frac{1}{2}\|{\textbf{\emph{y}}}-\boldsymbol\Psi\textbf{\emph{x}}\|_2^2+ \rho \boldsymbol\Theta(\textbf{\emph{x}}),}
\label{eq:2}
\end{equation} 
where $\|{\textbf{\emph{y}}}-\boldsymbol\Psi\textbf{\emph{x}}\|_2^2$ denotes the fidelity term for the inverse problem (\ie,  $\boldsymbol\Psi = \boldsymbol\Phi$ for image CS), and $\boldsymbol\Theta(\textbf{\emph{x}})$ denotes the prior based on certain deep denoiser \cite{zhang2017learning,zhang2018ffdnet,ulyanov2018deep}. $\rho$ is a regularization parameter. By applying highly effective deep priors, the deep PnP approaches have achieved superior results in many image processing applications \cite{zhang2017learning,8464091,8489894}.

{{\bf  Low-Rank Image Modeling:}
%
%
Besides deep image priors, other image properties such as NSS \cite{mairal2009non}, \ie, image patches are typically similar to other non-local structures within the same image, have been widely utilized for image restoration \cite{zhang2014group,mairal2009non,gu2017weighted,8931738}.
Popular NSS-based methods proposed to group the similar patches, and exploit the patch correlation within each group.
Specifically, $n$ overlapping patches $\textbf{\emph{x}}_i\in{\mathbb R}^{b}$ are extracted from the image $\textbf{\emph{x}}$.
Taking each $\textbf{\emph{x}}_i$ as the reference patch, its $m$ most similar patches are selected to  construct each data matrix $\textbf{\emph{X}}_i\in{\mathbb R}^{b \times m}$.

There are different methods to process the constructed data matrices, among which the low-rank (LR) modeling has demonstrated superior performance in many image restoration applications \cite{gu2017weighted,zhang2014group,geng2018truncated,8931738}. Comparing to PnP approaches based on deep prior \cite{zhang2017learning,8464091,8489894,ulyanov2018deep}, the LR-based methods are \textit{unsupervised} and typically limited by the model flexibility. Moreover, such methods inevitably produce the ringing artifacts due to the aggregation of overlapping patches.

\vspace{-2mm}
\section{Proposed Method}
\label{sec:4}

As mentioned above, NSS-based methods and deep learning based methods have their respective merits and drawbacks. In this section, we propose a general Hybrid PnP (H-PnP) framework by combining the two triply-complementary priors, dubbed the joint low-rank and deep (LRD) prior.
\vspace{-2mm}
\subsection{H-PnP Framework}
\label{sec:4.1}
We now propose a novel H-PnP framework for image CS by incorporating the LRD prior by solving the following optimization problem,
\begin{align}
&\textstyle{(\hat{{\textbf{\emph{x}}}}, \hat{{\textbf{\emph{L}}}}_i)= \argmin_{\textbf{\emph{x}}, \textbf{\emph{L}}_i}\frac{1}{2}\|{\textbf{\emph{y}}}- \boldsymbol\Phi\textbf{\emph{x}}\|_2^2 +\frac{\mu}{2}\sum\nolimits_{i=1}^n \|{\textbf{\emph{R}}}_i\textbf{\emph{x}}-{\textbf{\emph{L}}}_i\|_F^2}\nonumber \\
&\textstyle{\qquad \qquad \qquad +\lambda\sum\nolimits_{i=1}^n \textbf{\emph{P}} (\textbf{\emph{L}}_i)+\rho \boldsymbol\Theta(\textbf{\emph{x}}).}
\label{eq:5}
\end{align}
Similar to the deep PnP problem in Eq.~\eqref{eq:2}, the deep prior  $\boldsymbol\Theta(\textbf{\emph{x}})$ is applied in the proposed H-PnP scheme (\ie, Eq.~\eqref{eq:5}).  $\|\cdot\|_F^2$ denotes the Frobenius norm, and $\textbf{\emph{P}}(\cdot)$ is the low-rank regularizer with a non-negative weight $\lambda$. The rank penalties $\lambda\sum\nolimits_{i=1}^n \textbf{\emph{P}} (\textbf{\emph{L}}_i)$ are applied to exploit the image self-similarity. ${\textbf{\emph{R}}}_i \textbf{\emph{x}} = [{{\textbf{\emph{R}}}_{i}}_0 {\textbf{\emph{x}}, {\textbf{\emph{R}}}_{i}}_1 {\textbf{\emph{x}},\dots, {\textbf{\emph{R}}}_{i}}_{m-1} \textbf{\emph{x}} ]\in{\mathbb R}^{b \times m}$ denotes the matrix formed by the set of similar patches for each reference patch ${\textbf{\emph{x}}}_i$. The selected patches are vectorized and formed the columns of the matrix ${\textbf{\emph{R}}}_i \textbf{\emph{x}}$, which is approximated by a corresponding low-rank matrix ${\textbf{\emph{L}}}_i$.

The proposed H-PnP formulation incorporates the LRD prior, which assumes that the underlying image $\textbf{\emph{x}}$ satisfies both LR and deep priors. Comparing to traditional deep PnP problem of Eq.~\eqref{eq:2}, the proposed H-PnP scheme integrates  a pair of triply-{\em complementary priors} using one unified the optimization problem. 

\vspace{-2mm}
\section{Optimization for the Proposed Model}
\label{sec:4.2}
In this section, we present a highly effective image CS algorithm based on the proposed H-PnP using alternating minimizing. It can be seen that Eq.~\eqref{eq:5} is a large-scale non-convex optimization problem. To make the optimization tractable, we propose a simple alternating minimizing strategy to solve Eq.~\eqref{eq:5} for image CS, \ie, the algorithm alternates between solving Eq.~\eqref{eq:5} for  $\textbf{\emph{L}}_i$ and $\textbf{\emph{x}}$, which corresponds to the \emph{Low-rank Approximation} and \emph{Image Update} sub-problem, respectively.

\vspace{-2mm}
\subsection{Low-Rank Approximation}
\label{sec:4.2}
For fixed $\textbf{\emph{x}}$, we solve Eq.~\eqref{eq:5}  for each $\textbf{\emph{L}}_i$ by minimizing
\begin{equation}
\textstyle{\hat{{\textbf{\emph{L}}}}_i = \argmin_{\textbf{\emph{L}}_i}\frac{1}{2}\|{\textbf{\emph{R}}}_i\textbf{\emph{x}}-{\textbf{\emph{L}}}_i\|_F^2+\frac{\lambda}{\mu} \textbf{\emph{P}} (\textbf{\emph{L}}_i), \forall i = 1, ... , n.}
\label{eq:6}
\end{equation} 
In this work, we set the rank penalty $\textbf{\emph{P}} (\textbf{\emph{L}}_i)$ to be the weighted nuclear norm, as the corresponding WNNM method \cite{gu2017weighted} demonstrated superior performance in image restoration amongst other classical CS methods. There exists a closed-form solution to $\hat{{\textbf{\emph{L}}}}_i$ by applying the singular value decomposition (SVD), with details steps in  \cite{gu2017weighted}.

\vspace{-2mm}
\subsection{Image Update}
\label{sec:4.3}
For fixed low-rank matrix $\{\textbf{\emph{L}}_i\}_{i = 1}^{n}$, the image $\textbf{\emph{x}}$ can be updated by solving the following problem,
\begin{equation}
\textstyle{\hat{{\textbf{\emph{x}}}} = \argmin_{\textbf{\emph{x}}}\frac{1}{2}\|{\textbf{\emph{y}}}- \boldsymbol\Phi\textbf{\emph{x}}\|_2^2 +\frac{\mu}{2}\sum\nolimits_{i=1}^n \|{\textbf{\emph{R}}}_i\textbf{\emph{x}}-{\textbf{\emph{L}}}_i\|_F^2
+\rho \boldsymbol\Theta(\textbf{\emph{x}}).}
\label{eq:9}
\end{equation} 
One can observe that it is quite difficult to solve Eq.~\eqref{eq:9} directly.  In order to facilitate the optimization, we adopt the alternating direction method of multipliers (ADMM) algorithm \cite{boyd2011distributed} to solve ${{\textbf{\emph{x}}}}$.  Specifically, we  introduce an auxiliary variable $\textbf{\emph{z}}$ with the constraint $\textbf{\emph{x}} = \textbf{\emph{z}}$, then Eq.~\eqref{eq:9} can be rewritten as the following constrained problem,
\begin{align}
&\textstyle{(\hat{{\textbf{\emph{x}}}}, \hat{{\textbf{\emph{z}}}}) = \argmin_{\textbf{\emph{x}}, {\textbf{\emph{z}}}}\frac{1}{2}\|{\textbf{\emph{y}}}- \boldsymbol\Phi\textbf{\emph{x}}\|_2^2 +\frac{\mu}{2}\sum\nolimits_{i=1}^n \|{\textbf{\emph{R}}}_i\textbf{\emph{x}}-{\textbf{\emph{L}}}_i\|_F^2}\nonumber\\
&\textstyle{\qquad\qquad+\rho \boldsymbol\Theta(\textbf{\emph{z}}), ~~~{ s. t.} \quad \textbf{\emph{z}} = \textbf{\emph{x}}.}
\label{eq:10}
\end{align}
%
We then invoke ADMM algorithm by iterating the following variable updates Eq.~\eqref{eq:11} to Eq.~\eqref{eq:13},
\begin{align}
&\textstyle{\hat{{\textbf{\emph{x}}}} \leftarrow \argmin_{\textbf{\emph{x}}}\frac{1}{2}\|{\textbf{\emph{y}}}- \boldsymbol\Phi\textbf{\emph{x}}\|_2^2 +\frac{\mu}{2}\sum\nolimits_{i=1}^n \|{\textbf{\emph{R}}}_i\textbf{\emph{x}}-{\textbf{\emph{L}}}_i\|_F^2}\nonumber\\
&\textstyle{\qquad \qquad + \frac{\tau}{2}\|{\textbf{\emph{x}}}- {\textbf{\emph{z}}} -{\textbf{\emph{c}}} \|_2^2,}
\label{eq:11}\\
&\textstyle{\hat{{\textbf{\emph{z}}}} \leftarrow \argmin_{\textbf{\emph{z}}}\frac{\tau}{2}\|{\textbf{\emph{x}}} - {\textbf{\emph{z}}}-{\textbf{\emph{c}}} \|_2^2 +\rho \boldsymbol\Theta(\textbf{\emph{z}}),}
\label{eq:12}\\
&\textstyle{\hat{{\textbf{\emph{c}}}} \leftarrow {\textbf{\emph{c}}}  - ({\textbf{\emph{x}}}  - {\textbf{\emph{z}}} ),}
\label{eq:13}
\end{align} 
where $\tau$ is a balancing factor. One can observe that the sub-problems  Eq.~\eqref{eq:11} and  Eq.~\eqref{eq:12} for updating ${\textbf{\emph{x}}}$ and ${\textbf{\emph{z}}}$, respectively, have efficient solutions. We will introduce the corresponding details below.


\vspace{-2mm}
\subsubsection{ $\textbf{\emph{x}}$ \textbf{Sub-problem}}

Given the obtained $\{{\textbf{\emph{L}}}_i\}_{i =1}^n$ and ${\textbf{\emph{z}}}$, $\textbf{\emph{x}}$ sub-problem in Eq.~\eqref{eq:11} is essentially a minimization problem of a strictly convex quadratic function. However, $\boldsymbol\Phi$ is a $M \times N$ random projection matrix without a specific structure in image CS, which is expensive to directly compute matrix inversion. To avoid this issue, a gradient descent method \cite{deift1993steepest} is applied, \ie,
\begin{equation}
\textstyle{\hat{\textbf{\emph{x}}} = \textbf{\emph{x}} - \eta\textbf{\emph{q}},}
\label{eq:15}
\end{equation}
where $\eta$ represents the step size, and $\textbf{\emph{q}}$ is the gradient of the objective, which can be calculated as
\begin{equation}
\begin{aligned}
\textstyle{\hat{\textbf{\emph{x}}}} &= \textstyle{\textbf{\emph{x}} - \eta \left[ \boldsymbol\Phi^T\boldsymbol\Phi\textbf{\emph{x}}- \boldsymbol\Phi^T \textbf{\emph{y}} +\tau (\textbf{\emph{x}} - \textbf{\emph{z}}- \textbf{\emph{c}}) \right.}\\
&\textstyle{\qquad\left.+ \mu(\sum\nolimits_{i=1}^n {\textbf{\emph{R}}}_i^T{\textbf{\emph{R}}}_i\textbf{\emph{x}}-\sum\nolimits_{i=1}^n {\textbf{\emph{R}}}_i^T{\textbf{\emph{R}}}_i\textbf{\emph{L}}_i)\right],}
\label{eq:16}
\end{aligned}
\end{equation}
where both $\boldsymbol\Phi^T\boldsymbol\Phi$ and $\boldsymbol\Phi^T\textbf{\emph{y}}$ are pre-computed and fixed during the iterations.

\vspace{-2mm}
\begin{center}
\begin{algorithm}[!t]
\caption{The Proposed H-PnP Algorithm for Image CS.}
\begin{algorithmic}[1]
\REQUIRE Measurement $\textbf{\emph{y}}$ and the sensing matrix $\boldsymbol\Phi$.
\STATE Set parameters $b$, $m$, $W$, $K$, $\mu$, $\rho$, $\tau$,  $\lambda$, $\upsilon$ and $\eta$.
\STATE \textbf{Initialization:} Estimate an initial image $\hat{\textbf{\emph{x}}}$ using a standard CS method (\eg, DCT/MH \cite{chen2011compressed}  based reconstruction method).
\FOR{$k = 1, \dots, K$ }
\STATE Divide $\hat{\textbf{\emph{x}}}^{k}$ into a set of overlapping patches with size $\sqrt{b}\times\sqrt{b}$;
\vspace{-4mm}
\FOR{Each patch ${\textbf{\emph{x}}}_{i}$ in ${\textbf{\emph{x}}}^k$ }		
\STATE Find similar patches to form a group matrix ${\textbf{\emph{X}}}_{i}$;
\STATE Singular value decomposition (SVD) for  ${\textbf{\emph{X}}}_{i}$;
\STATE Update $\hat{\textbf{\emph{L}}}_i$ by invoking WNNM \cite{gu2017weighted};
\ENDFOR
\STATE \textbf{ADMM:}
\STATE  Initialization: $\textbf{\emph{c}} = 0$;  $\textbf{\emph{z}} = \hat{\textbf{\emph{x}}}^{k}$;
\STATE Update $\hat{\textbf{\emph{x}}}$ by computing Eq.~\eqref{eq:16};
\STATE Update $\hat{\textbf{\emph{z}}}$ by computing Eq.~\eqref{eq:18};
\STATE Update $\hat{\textbf{\emph{c}}}$ by computing Eq.~\eqref{eq:13};
\ENDFOR
\STATE $\textbf{Output:}$ The final restored image $\hat{\textbf{\emph{x}}}$ = $\textbf{\emph{x}}^{(k+1)}$.
\end{algorithmic}
\label{algo:1}
\end{algorithm}
\end{center}
\vspace{-4mm}
\subsubsection{ $\textbf{\emph{z}}$ \textbf{Sub-problem}}

Given $\textbf{\emph{x}}$, then $\textbf{\emph{z}}$ sub-problem in Eq.~\eqref{eq:12}  can be rewritten as
\begin{equation}
\textstyle{\min_{\textbf{\emph{z}}}\textbf{\emph{Q}}_2(\textbf{\emph{z}}) = \min_{\textbf{\emph{z}}}\frac{1}{2(\sqrt{\rho/\tau})^2}\|{\textbf{\emph{r}}} - {\textbf{\emph{z}}} \|_2^2 + \boldsymbol\Theta(\textbf{\emph{z}}),}
\label{eq:17}
\end{equation} 
where ${\textbf{\emph{r}}} = {\textbf{\emph{x}}}- {\textbf{\emph{c}}}$. From a Bayesian perspective, Eq.~\eqref{eq:17} is a Gaussian denoising problem for ${\textbf{\emph{z}}}$ by solving a MAP problem, with the corresponding  noise standard deviation to be $\sqrt{\rho/\tau}$ \cite{zhang2017learning}. Accordingly, we denote such denoising problem as 
\begin{equation}
\textstyle{\hat{{\textbf{\emph{z}}}} = \cal{F} (\textbf{\emph{r}}, \sqrt{\rho/\tau}),}
\label{eq:18}
\end{equation} 
where $\cal{F}$ denotes a Gaussian denoiser based on the specific deep image prior $\boldsymbol\Theta(\cdot)$. In general, any deep image Gaussian denoisier can be used as the $\cal{F}(\cdot)$ in Eq.~\eqref{eq:18}. In this paper, we apply a fast and flexible denoising CNN (FFDNet) \cite{zhang2018ffdnet} for Eq.~\eqref{eq:18}, which is an efficient but effective CNN-based denoisier. Furthermore, FFDNet is capable of dealing with different standard deviations by self-adaption.

Till now, the efficient solution for each separated minimization sub-problem has been achieved, which makes the whole algorithm efficient and effective. After solving the above sub-problems, the complete description of the proposed H-PnP algorithm for image CS is summarized in Algorithm~\ref{algo:1}. The iterative algorithm in this paper is terminated when ${||{\hat{\textbf{\emph{x}}}}^{k}-{\hat{\textbf{\emph{x}}}}^{k-1}||_2^2}/{||{\hat{\textbf{\emph{x}}}}^{k-1}||_2^2}<\upsilon$, where $\upsilon$ is a small constant.

\vspace{-2mm}
\section{Experimental Results}
\label{sec:5}
We conduct extensive experiments to evaluate the proposed H-PnP algorithm for image CS.  Similar to the setups in previous works \cite{shi2019scalable,zhang2014group,li2009user,chen2011compressed}, we simulate the image CS measurements at the block level (with the block size of $32\times32$) using a Gaussian random projection matrix for each test image. The image CS algorithms are applied to reconstruct the image using the simulated CS measurements. We apply the peak signal to noise ratio (PSNR) to evaluate the CS reconstructed images. 

\vspace{-2mm}
\subsection{Implementation and Parameters}
\label{sec:5.1}

We applied the pre-trained FFDNet denoiser \cite{zhang2018ffdnet} as the deep prior in the proposed H-PnP based image CS algorithm. The main parameters of the proposed method are set as follows. The size of each patch $\sqrt{b} \times \sqrt{b}$ is set to 7$\times$7, the number of patches grouped by KNN operator is $m$ = 60, and the size of KNN search window $W \times W$ is set to 20$\times$20. We set the maximum number of iterations to be $K$ = 60. The coefficients $\mu$, $\lambda$, $\rho$ and $\tau$ are tuned for different sampling ratio in image CS (see the tuning procedure as well as the values for each sampling ratio in our shared code). The source code of the proposed HPnP method for image CS is available at: \url{https://drive.google.com/open?id=1HcgKtj0r5SVpp6yKmRVyI9b8TdXjcT83}.


\begin{table}[!t]
\vspace{-4mm}
\caption{ \footnotesize Average PSNR ($\textnormal{d}$B) comparison of TV \cite{li2009user}, Rcos \cite{zhang2012image}, GSR \cite{zhang2014group}, JASR \cite{eslahi2016compressive}, TNNM \cite{geng2018truncated}, WNNM \cite{gu2017weighted} and the proposed H-PnP on BSD68 dataset \cite{arbelaez2010contour}.}
\footnotesize
\centering
\resizebox{0.48\textwidth}{!}
{
\begin{tabular}{|c|c|c|c|c|c|c|c|c|c|c|}
\hline
\textbf{Methods} & \multicolumn{1}{c|}{0.1}  & \multicolumn{1}{c|}{0.2}& \multicolumn{1}{c|}{0.3}
& \multicolumn{1}{c|}{0.4}& \multicolumn{1}{c|}{0.5}& \multicolumn{1}{c|}{\textbf{Average}}\\
\hline
\multirow{1}{*}{TV \cite{li2009user}}
&	24.93 	&	27.31 	&	29.15 	&	30.88 	&	32.58 	&	28.97
\\
\hline
\multirow{1}{*}{Rcos \cite{zhang2012image}}
&	26.03 	&	28.68 	&	30.62 	&	32.33 	&	34.03 	&	30.34
\\
\hline
\multirow{1}{*}{GSR \cite{zhang2014group}}
&	25.83 	&	29.28 	&	31.82 	&	34.02 	&	36.11 	&	31.41
\\
\hline
\multirow{1}{*}{JASR \cite{eslahi2016compressive}}
&	26.19 	&	29.46 	&	31.63 	&	33.52 	&	35.33 	&	31.22
\\
\hline
\multirow{1}{*}{TNNM \cite{geng2018truncated}}
&	26.52 	&	29.93 	&	32.31 	&	34.36 	&	36.31 	&	31.88
\\
\hline
\multirow{1}{*}{WNNM \cite{gu2017weighted}}
&	26.60 	&	29.84 	&	32.28 	&	34.43 	&	36.54 	&	31.94
\\
\hline
\multirow{1}{*}{\textbf{Proposed}}
&	\textbf{27.41} 	&	\textbf{30.49} 	&	\textbf{32.79} 	&	\textbf{34.86} 	&	\textbf{36.82} 	&	\textbf{32.47}
\\
\hline
\end{tabular}}
\label{Tab:2}
\vspace{-4mm}
\end{table}

\begin{figure}[!htbp]
\centering
\vspace{-2mm}
{\includegraphics[width= 0.48\textwidth]{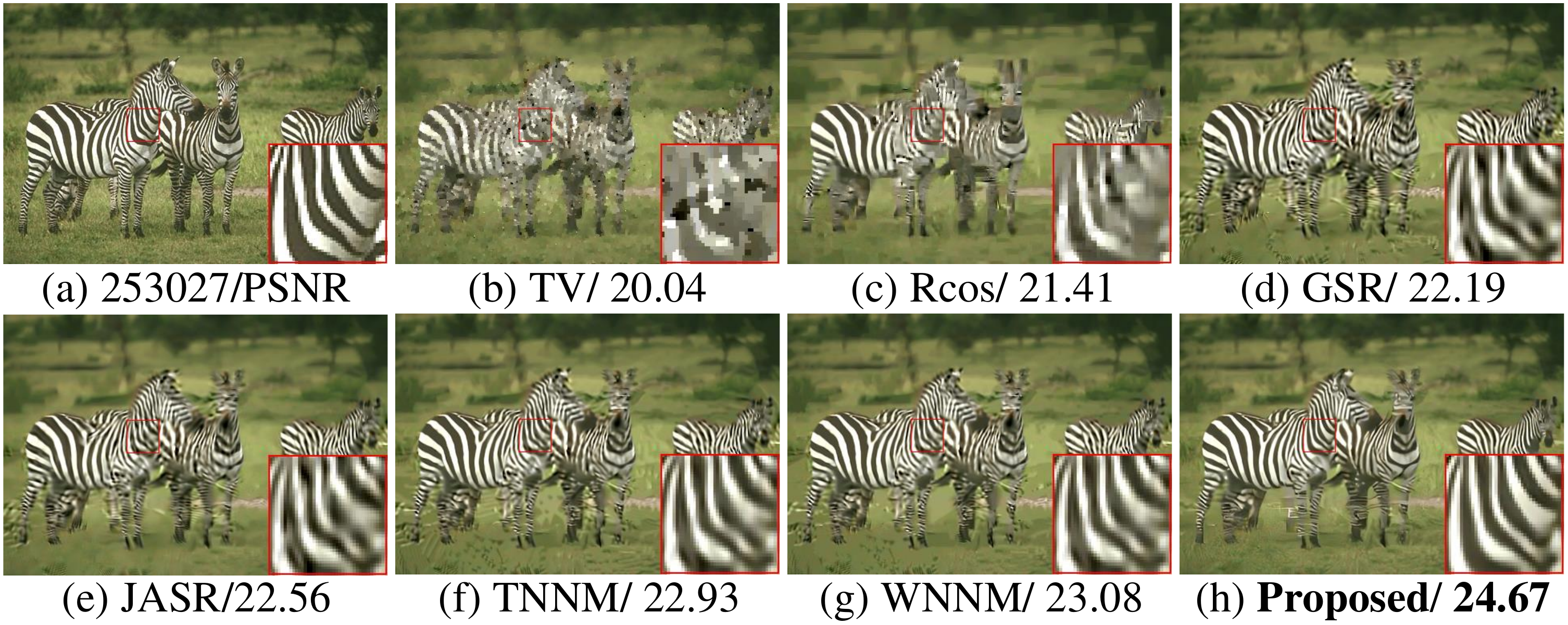}}
\vspace{-6mm}
\caption{\footnotesize Visual quality comparisons of image CS on image $\emph{253027}$ from BSD68 BSD68 \cite{arbelaez2010contour}  in the case of sampling ratio = 0.1.}
\label{fig:2}
\vspace{-2mm}
\end{figure}

\vspace{-2mm}
\subsection{Comparison with Classical Image CS Methods}
\label{sec:5.1}
We first compare the proposed H-PnP image CS algorithm to the popular or state-of-the-art classic methods, \ie, TV \cite{li2009user}, Rcos \cite{zhang2012image}, GSR \cite{zhang2014group}, JASR \cite{eslahi2016compressive}, TNNM \cite{geng2018truncated} and WNNM \cite{gu2017weighted}. Amongst them, WNNM is a well-known non-local method which provides the state-of-the-art results for image denoising. We extend the WNNM denoising algorithm \cite{gu2017weighted} to image CS by applying ADMM \cite{boyd2011distributed} which is similar to what we described in Section~\ref{sec:4.3}. The extended WNMM for image CS achieves the best results among all classic competing methods. We use the publicly available codes of other competing methods from their official websites with the default parameter settings for all experiments.

We simulate the image CS measurements for all test images using five different sampling ratios, \ie, 0.1, 0.2, 0.3, 0.4 and 0.5. Table~\ref{Tab:2} lists the average PSNRs  over all test images from BSD68 (68 images) \cite{arbelaez2010contour}, obtained by our proposed H-PnP based image CS algorithm, as well as the six classic competing methods. It is clear that our proposed H-PnP  consistently outperforms all competing methods on different sampling ratios for all datasets. On average, our proposed H-PnP enjoys a PSNR gain over TV by 3.50dB, over Rcos by 2.14dB, over GSR by 1.06dB, over JASR by 1.25dB, over TNNM by 0.59dB and over WNNM by 0.54dB. The visual quality comparisons of image \emph{253027} in the case of sampling ratio of 0.1 are shown in Fig.~\ref{fig:2}. We have magnified a sub-region of each image to compare visual result of each competing method. It can be seen that TV cannot obtain a visual pleasant result. Rcos, GSR, JASR, TNNM and WNNM methods are all prone to produce some undesirable ringing artifacts. By contrast, the proposed H-PnP algorithm not only preserves fine image details, but also removes the visual artifacts significantly.

\vspace{-2mm}
\subsection{Comparison with Deep Image CS Methods}
\label{sec:5.2}
We now compare our proposed H-PnP with deep learning based methods including: SDA \cite{mousavi2015deep}, ReconNet \cite{kulkarni2016reconnet}, IST-Net \cite{zhang2018ista}, IST-Net$^{+}$ \cite{zhang2018ista}, CSNet \cite{8765626} and SCSNet \cite{shi2019scalable} methods. Note that SCSNet exploited a scale CNN that delivers state-of-the-art image CS  performance. We follow \cite{zhang2018ista} to use the images on BSD68 \cite{arbelaez2010contour} as the test images. The average PSNR results of our proposed H-PnP as well as different deep learning based methods on four sampling ratios are shown in Table~\ref{Tab:3}, where the results of SDA, ReconNet, ISTA-Net and ISTA-Net$^{+}$ are from  \cite{zhang2018ista}. One can observe that our proposed H-PnP significantly outperforms all deep learning based methods. In particular, the proposed H-PnP achieves 0.76dB gains in average PSNR over SCSNet method. The visual comparisons of image $\emph{119082}$ on BSD68 dataset are shown in Fig.~\ref{fig:4}. It can be observed that some visual artifacts are still visible in all competing deep learning based methods.  By contrast, our proposed H-PnP not only significantly removes undesirable artifacts across all the image, but also preserves large-scale sharp edges and small-scale fine details.

\begin{table}[!t]
\vspace{-4mm}
\caption{ \footnotesize Average PSNR ($\textnormal{d}$B) comparison of different deep learning based image CS reconstruction methods on BSD68 dataset \cite{arbelaez2010contour}.}
\tiny
\centering
\resizebox{0.48\textwidth}{!}
{
\begin{tabular}{|c|c|c|c|c|c|c|c|c|c|}
\hline
\textbf{Methods} & \multicolumn{1}{c|}{0.1}  & \multicolumn{1}{c|}{0.3}
& \multicolumn{1}{c|}{0.4}& \multicolumn{1}{c|}{0.5}& \multicolumn{1}{c|}{Average}\\
\hline
\multirow{1}{*}{{SDA} \cite{mousavi2015deep}}
&	23.12 	&	26.38 	&	27.41 	&	28.35 	&	26.32
\\
\hline
\multirow{1}{*}{{ReconNet} \cite{kulkarni2016reconnet}}
&	24.15 	&	27.53 	&	29.08 	&	29.86 	&	27.66
\\
\hline
\multirow{1}{*}{{IST-Net} \cite{zhang2018ista}}
&	25.02 	&	29.93 	&	31.85 	&	33.60 	&	30.10
\\
\hline
\multirow{1}{*}{{IST-Net$^{+}$} \cite{zhang2018ista}}
&	25.33 	&	30.34 	&	32.21 	&	34.01 	&	30.47
\\
\hline
\multirow{1}{*}{{CSNet} \cite{8765626}}
&	27.10 	&	31.45 	&	33.46 	&	34.90 	&	31.73
\\
\hline
\multirow{1}{*}{{SCSNet} \cite{shi2019scalable}}
&	27.28 	&	31.88 	&	33.87 	&	35.79 	&	32.21
\\
\hline
\multirow{1}{*}{\textbf{Proposed}}
&	\textbf{27.41} 	&	\textbf{32.79} 	&	\textbf{34.86} 	&	\textbf{36.82} 	&	\textbf{32.97}
\\
\hline
\end{tabular}}
\label{Tab:3}
\vspace{-4mm}
\end{table}

\begin{figure}[!htbp]
\centering
	\vspace{-2mm}
\begin{minipage}[b]{1\linewidth}
{\includegraphics[width= 1\textwidth]{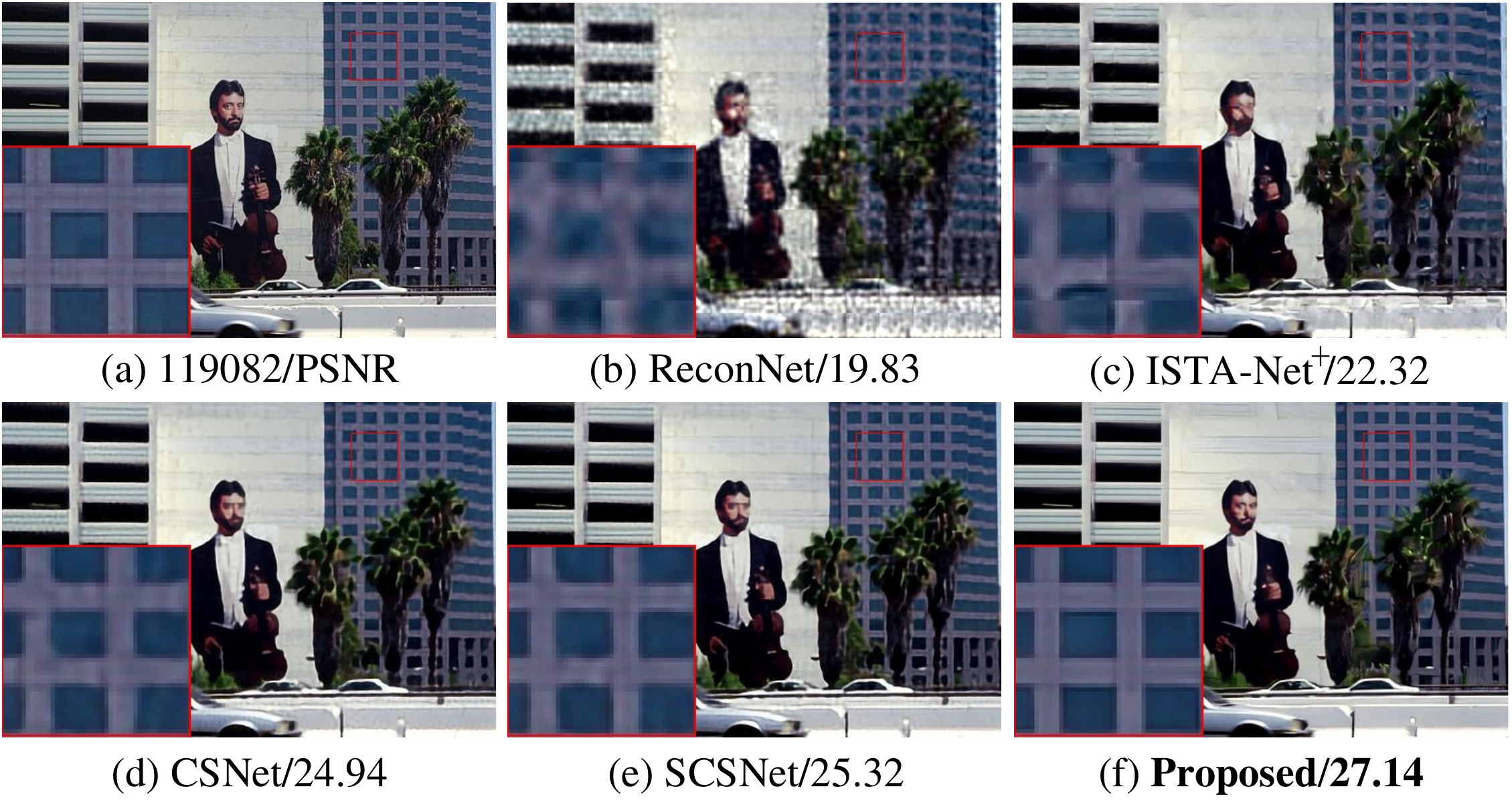}}
\end{minipage}
\vspace{-6mm}
\caption{\footnotesize Visual quality comparisons of image CS on image $\emph{119082}$ from BSD68 \cite{arbelaez2010contour} in the case of sampling ratio = 0.1.}
\label{fig:4}
\vspace{-4mm}
\end{figure}

\vspace{-2mm}
\section{Conclusion}
\label{sec:6}

This paper proposed a joint low-rank and deep (LRD) image model, which comprises a pair of triply complementary priors, namely \textit{external} and \textit{internal}, \textit{deep} and \textit{shallow}, and \textit{local} and \textit{non-local} priors. We have then proposed a H-PnP framework based on the LRD model to solve image CS problem along with an alternating minimization method. Experimental results have demonstrated that the proposed H-PnP based image CS algorithm significantly outperforms  many state-of-the-art image CS methods. Future work lies in the theoretical analysis of our proposed model and apply it to other image restoration tasks.

{\footnotesize
\bibliographystyle{IEEEbib}
\bibliography{lrd1}
}

\end{document}